\documentclass[9pt,twocolumn,twoside]{opticajnl}
\journal{opticajournal} 
\setboolean{shortarticle}{true}
\usepackage{lineno}
\usepackage{upgreek}
\usepackage{scalerel}
\usepackage{tikz}
\usetikzlibrary{svg.path,calc}

\definecolor{orcidlogocol}{HTML}{A6CE39}
\tikzset{
   orcidlogo/.pic={
    \fill[orcidlogocol] svg{M256,128c0,70.7-57.3,128-128,128C57.3,256,0,198.7,0,128C0,57.3,57.3,0,128,0C198.7,0,256,57.3,256,128z};
    \fill[white] svg{M86.3,186.2H70.9V79.1h15.4v48.4V186.2z}
                 svg{M108.9,79.1h41.6c39.6,0,57,28.3,57,53.6c0,27.5-21.5,53.6-56.8,53.6h-41.8V79.1z M124.3,172.4h24.5c34.9,0,42.9-26.5,42.9-39.7c0-21.5-13.7-39.7-43.7-39.7h-23.7V172.4z}
                 svg{M88.7,56.8c0,5.5-4.5,10.1-10.1,10.1c-5.6,0-10.1-4.6-10.1-10.1c0-5.6,4.5-10.1,10.1-10.1C84.2,46.7,88.7,51.3,88.7,56.8z};
  }
}

\newcommand\orcidicon[1]{\href{https://orcid.org/#1}{\mbox{
\begin{tikzpicture}[overlay,remember picture]
\coordinate (A);
\coordinate(B) at ($(A)-(2pt,-9pt)$);
\end{tikzpicture}
\begin{tikzpicture}[overlay,remember picture,yscale=-0.0381,xscale=0.0381,transform shape]
\pic at (B) {orcidlogo};
\end{tikzpicture}
}{}}}

\usepackage{hyperref}
\dates{}   
\doi{}

\title{Ultra-low-noise supercontinuum in normal-dispersion ZBLAN fibres pumped at 1.85~$\upmu$m}
\author[1,2,*,\protect\orcidicon{0000-0003-2007-0930}\,\,]{ Shreesha Rao D. S.}
\author[2,\protect\orcidicon{0000-0002-9302-6124}\,\,]{Anupamaa Rampur}
\author[1,3,4,\protect\orcidicon{0000-0002-8041-9156}\,\,]{Ole Bang}
\author[2,\protect\orcidicon{0000-0001-9073-2894}\,\,]{Alexander M. Heidt}

\affil[1]{DTU Electro, Dept. of Electrical and Photonics Engineering, Technical University of Denmark, {\O}rsteds Plads, 2800 Kongens Lyngby, Denmark}
\affil[2]{Institute of Applied Physics, University of Bern, Sidlerstrasse 5, 3012 Bern, Switzerland}
\affil[3]{NKT Photonics A/S, Blokken 84, 3460 Birker{\o}d, Denmark}
\affil[4]{NORBLIS ApS, {\O}rsteds Plads 343, 2800 Kongens Lyngby, Denmark}
\affil[*]{sheds@dtu.dk}
\begin{abstract}
We demonstrate, for the first time to our knowledge, ultra-low-noise supercontinuum (SC) generation in normal-dispersion fluoride fibres pumped by femtosecond (fs) pulses. We have investigated two elliptical-core polarisation-maintaining (PM) ZBLAN fibres with core dimensions 6.7$\times$2.7~$\upmu$m and 8.9$\times$4.1~$\upmu$m, experimentally measured to have normal dispersion up to 3.77~$\upmu$m and 3.25~$\upmu$m, respectively; the smaller-core fibre yields ultra-low-noise SC spanning 1.537--2.196~$\upmu$m with a minimum relative-intensity noise (RIN) of 0.22\% at 1.7~$\upmu$m, and the larger-core fibre yields 1.507--2.250~$\upmu$m with 0.36\% at 2.0~$\upmu$m. To aid the generation of low-noise SC, we developed an all-PM thulium chirped-pulse amplifier delivering 58~fs pulses at 1.85~$\upmu$m, 210~mW average power at 40~MHz, with 0.41\% RIN, seeded by a part of an ultra-low-noise SC using a 1.55~$\upmu$m fs laser and an all-normal-dispersion (ANDi) silica fibre for precise seed control. These results establish a robust, alignment-free pathway to extend ultra-low-noise ANDi-fibre SC towards the mid-infrared using PM fluoride fibres.
\end{abstract}
\setboolean{displaycopyright}{false} 
\begin{document}
\maketitle
Ultra-low-noise supercontinuum (SC) can be generated when spectral broadening arises from the coherent nonlinear processes of self-phase modulation and optical wave-breaking~\cite{Het10FltTp}. In practice, spectral broadening can be achieved through these processes when a femtosecond (fs) pulse is launched into a fibre with normal-dispersion~\cite{Hei17LimJb}. To preserve ultra-low-noise characteristics, it must be ensured that the generated spectrum is entirely within the normal-dispersion region of the fibre~\cite{Ull19ShrOl}. Such SC sources can exhibit a smooth spectral profile along with a clean single-pulse temporal structure~\cite{Fin08BeIJb}.

SC generation (SCG) has been studied in silica-based speciality all-normal dispersion (ANDi) fibres using various fs pump sources---for example, around 0.8~$\upmu$m~\cite{Hei11JosabCoh,Hoop11IeeCoh}, 1~$\upmu$m~\cite{Wan07Oc1um,Liu151umJlt}, 1.55~$\upmu$m~\cite{Nis041o5AOl,Cho06Pm15El}, and 1.75~$\upmu$m~\cite{Nis07JosaB1o7,Tar17Oe1o7}. ANDi fibre-based fs-pumped ultra-low-noise sources have enabled a range of applications in the near-infrared (NIR), including high-quality single-pulse generation~\cite{Dem11Oe1o3c}, shot-noise-limited ultra-high-resolution (UHR) optical coherence tomography (OCT)~\cite{Shr21StOctLsa}, UHR scanning near-field optical microscopy (SNOM)~\cite{Kob21SnomApl}, and dual-comb spectroscopy~\cite{Gru24DuaOl}, among others.

One of the limiting factors that can degrade the noise properties of an fs-pumped SC in low-birefringence ANDi fibres is polarisation mode instability (PMI)~\cite{Mil98NpmiJb,SrPmi18Ivn}. However, PMI can be completely eliminated using high-birefringence fibres~\cite{Liu151umJlt,Eti20XpmiOl,Rao22PmNpm}, also known as polarisation-maintaining (PM) fibres. Although fs-pumped, ANDi fibre-based ultra-low-noise SC sources covering the entire transmission window of silica fibre have not been achieved, SC sources spanning various wavelength bands have been extensively studied, as noted above. In contrast, fs-pumped SCG in ANDi fibres within the mid-infrared (MIR) region is only beginning to be explored.

Normal-dispersion fibre and fs-pumped SCG in the MIR have been demonstrated in tellurite~\cite{Pau18TelAs,Klim19TelJb} and chalcogenide~\cite{Wan17AsS,Xin18ChacOe,Zha19TeChOe} fibres. Extending ANDi fibre-based SCG into the MIR requires both suitable fibre designs and appropriate fs pumps. Spectra extending beyond 2.7~$\upmu$m~\cite{Wan17AsS,Zha19TeChOe} have been demonstrated using pumps operating at wavelengths longer than 3~$\upmu$m, typically, tunable fs laser systems such as optical parametric amplifiers. On the other hand, fluoride-based soft-glass fibres made from ZBLAN have been employed for SCG in the anomalous-dispersion regime~\cite{Hag06ZbScIe,Xia06ZbScOl}, as well as in other nonlinear optical studies~\cite{Gao13ElZbOl}. However, SCG in normal-dispersion fluoride fibres has not yet been demonstrated, primarily due to the unavailability of such fibres. Extending ultra-low-noise SCG into the MIR is critical for several emerging applications. For instance, it could substantially enhance the sensitivity of MIR OCT systems~\cite{Nie19MOctLsc,Shr21StOctLsa} and enable SNOM in the MIR---a capability currently unattainable due to the absence of suitable ultra-low-noise sources.

In this Letter, we report the development of an all-PM silica fibre-based laser delivering 58~fs pulses centred at 1.85~$\upmu$m. Using this source, we have investigated ultra-low-noise SCG in two normal-dispersion, elliptical-core PM ZBLAN fibres. To the best of our knowledge, this is the first demonstration of ultra-low-noise SCG in a normal-dispersion fluoride fibre.

We chose a chirped-pulse amplification (CPA) architecture to generate low-noise pump pulses for the SCG, as it provides good control over the pump pulse parameters. Specifically, we use a coherently broadened seed from a robust 1.56~$\upmu$m erbium (Er)-doped fs source and power scaling in a thulium-doped fibre amplifier. Whereas transfer to 2~$\upmu$m in many such systems relies on anomalous-dispersion broadening~\cite{Ime052umSOe} and can introduce pulse-to-pulse fluctuations~\cite{Cor03Noi1Prl,New03Noi2Ol}, we instead generate the seed coherently in a PM ANDi fibre. Together with the CPA layout, this enables independent control of seed generation, stretching, amplification, and compression, and provides stable, low-noise pump pulses at 1.85~$\upmu$m. Although directly mode-locked thulium fibre lasers~\cite{Nel952um1Apl,Sha962um2Ol} are also convenient pump sources, they generally offer less flexibility because pulse formation dynamics, chirp, noise, and output spectrum are more tightly coupled.

\begin{figure}[t!]
\centering
\fbox{\includegraphics[width=0.97\linewidth]{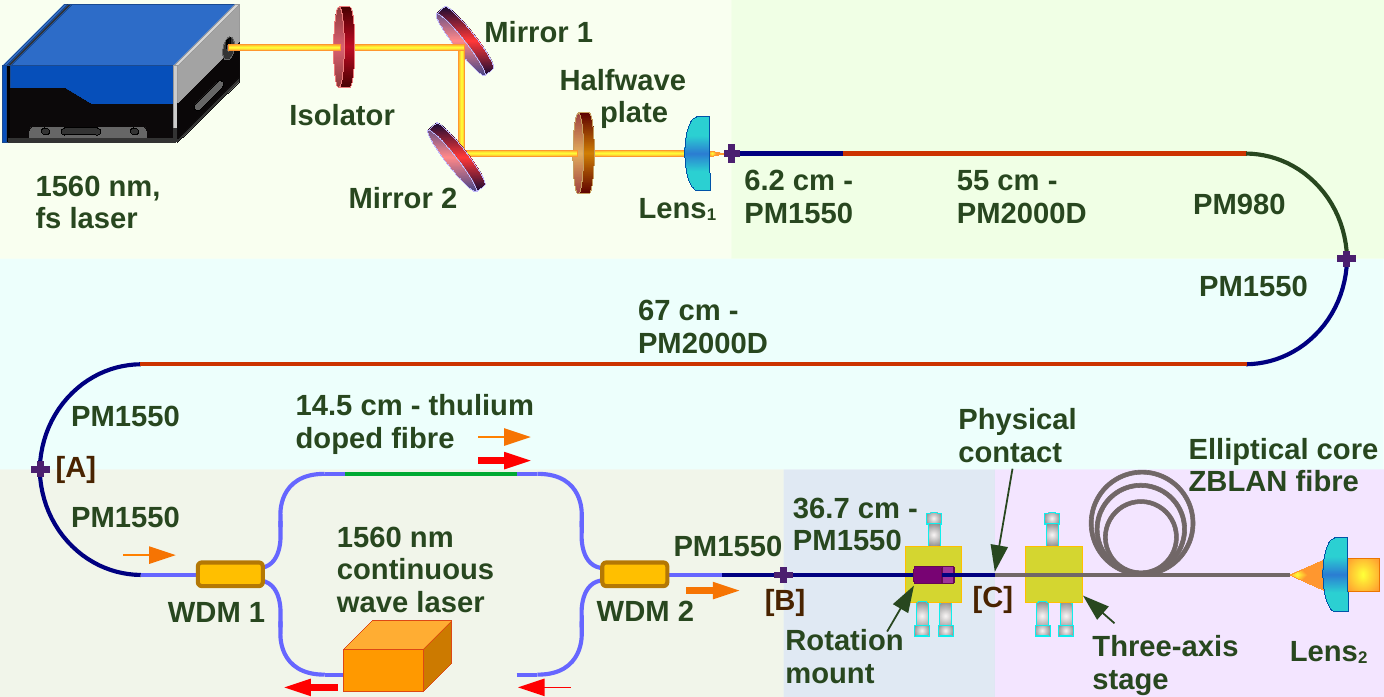}}
\caption{Schematic of the 1.85~$\upmu$m fs thulium amplifier and PM ZBLAN fibre in an all-PM architecture.}
\label{Fig:Las2um}
\end{figure}

\begin{figure}[htbp!]
\centering
\fbox{\includegraphics[width=0.97\linewidth]{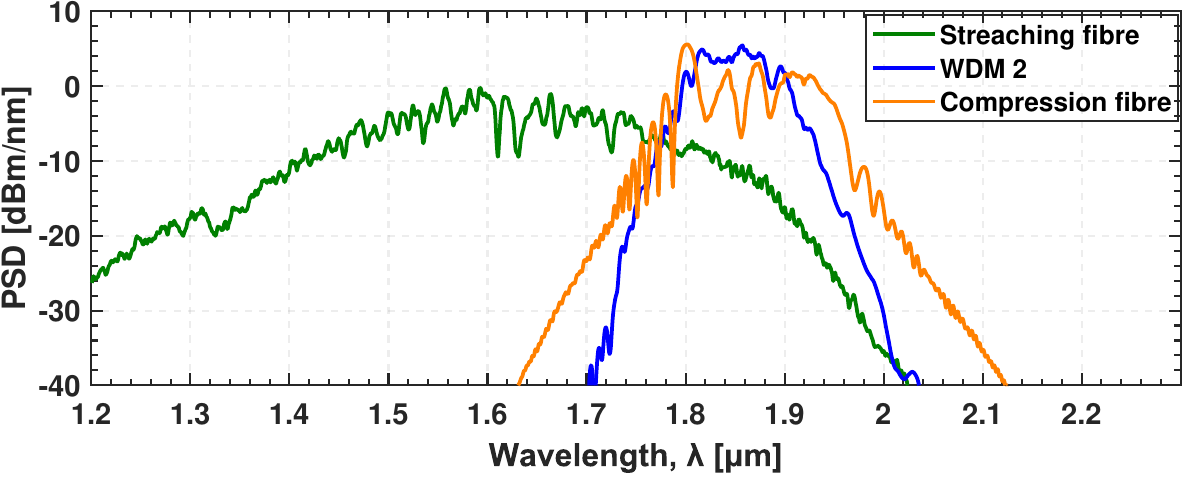}}
\caption{Experimental spectra at three locations: SC after the stretching fibre in green; amplified 1.85~$\upmu$m output after WDM~2 in blue; final compressed 1.85~$\upmu$m output in orange.}
\label{Fig:IntSpec}
\end{figure}

The schematic of the coherently seeded 1.85~$\upmu$m fs thulium CPA system is shown in Fig.~\ref{Fig:Las2um}. To effectively describe the system, it is divided into five sections. \emph{Section~I---1.56~$\upmu$m fs source:} A 1.56~$\upmu$m mode-locked fibre laser (Toptica: FemtoFibre pro) delivers 110~fs pulses with an average power of 320~mW at a repetition rate of 40~MHz. The output passes through an isolator (Thorlabs: IO-5-1550-HP), two low-group-delay-dispersion (GDD) mirrors (Thorlabs: UM10-45C), and a half-wave plate (Thorlabs: WPH10M-1550) before coupling through a collimator with an aspheric lens (Thorlabs: F220APC-1550) into the fibre. \emph{Section~II---ANDi fibre-based ultra-low-noise SCG:} We use a hybrid-ANDi fibre concept for cascaded nonlinear compression and SCG as described in Ref.~\cite{Ben22CasAOp}. The hybrid fibre consists of 6.2~cm of PM1550 fibre (Coherent) spliced to 55~cm of PM2000D (Coherent) ANDi fibre. The 110~fs input pulse undergoes soliton compression in the PM1550 section to $\sim$16~fs. This pulse generates a broadband, ultra-low-noise SC extending beyond 2~$\upmu$m in the ANDi fibre. PM1550 exhibits anomalous dispersion for wavelengths longer than 1.34~$\upmu$m, whereas PM2000D is ANDi. Their measured dispersion profiles are shown in Ref.~\cite{Ben22CasAOp}. The PM2000D output is spliced to a short PM980 (Thorlabs) fibre with FC/APC termination for connectorisation.As the SC is generated in the ANDi regime, the output remains a single chirped temporal pulse with a full-width at half-maximum (FWHM) duration of about 2~ps, and $<$500~W peak power. \emph{Section~III---pulse stretching:} The SC pulse is further dispersively stretched in 67~cm of PM2000D fibre to $>$9~ps FWHM. A simulated $B$-integral of the order of $10^{-4}$ confirms negligible nonlinear reshaping in this section. The power spectral density (PSD) after the stretching fibre, labelled [A] in Fig.~\ref{Fig:Las2um}, is shown in Fig.~\ref{Fig:IntSpec} in green. The spectra were measured using an optical spectrum analyser (OSA; Yokogawa: AQ6375).

\emph{Section~IV---thulium-doped amplification:} The stretched SC is launched into a wavelength division multiplexer (WDM)~1 (Opneti: PMIWDM-T2000/R1550), which combines a 1.56~$\upmu$m continuous-wave pump and the 1.85~$\upmu$m part of the ANDi SC for amplification in 14.5~cm of PM thulium-doped fibre (Coherent: PM-TDF-10P/130-HE). The thulium-doped fibre amplifier raises the stretched seed power in the 1.7--2.0~$\upmu$m range from $\sim$10 mW to 210 mW. WDM~1 acts as a long-pass filter with a cut-off near 1.62~$\upmu$m, below the useful thulium amplification band. WDM~2 separates the residual 1.56~$\upmu$m pump from the amplified signal centred at 1.85~$\upmu$m. The amplified spectrum, labelled [B] in Fig.~\ref{Fig:Las2um}, is shown in Fig.~\ref{Fig:IntSpec} in blue; the measured pulse duration is about 450~fs at this point. All WDM arms are made of PM1550 fibre. The input arm of WDM~1 and the output arm of WDM~2 are spliced to PM1550 fibre with FC/APC termination. \emph{Section~V---pulse compression:} The amplified pulse is subsequently compressed in 36.7~cm of PM1550 fibre. The length was determined by cutback to compensate for the accumulated GDD, thereby yielding the shortest pulse duration of 58~fs. The input side of the PM1550 fibre for compression is FC/APC terminated for connectorisation. The final output after compression, labelled [C] in Fig.~\ref{Fig:Las2um}, is shown in Fig.~\ref{Fig:IntSpec} in orange.

Sections~II--IV include short PM1550 pigtails with FC/APC termination to allow connectorisation, section replacement, and length optimisation during the development of the system. An exception is the output of Section~II, where PM980 fibre was used instead of PM1550 to enable proper characterisation of the SC and avoid distortion from multimode beating of short-wavelength SC components below 1.32~$\upmu$m, corresponding to the single-mode cut-off of PM1550; this does not affect the subsequent amplifier because it operates at 1.85~$\upmu$m. 
\begin{figure}[htbp!]
\centering
\fbox{\includegraphics[width=0.97\linewidth]{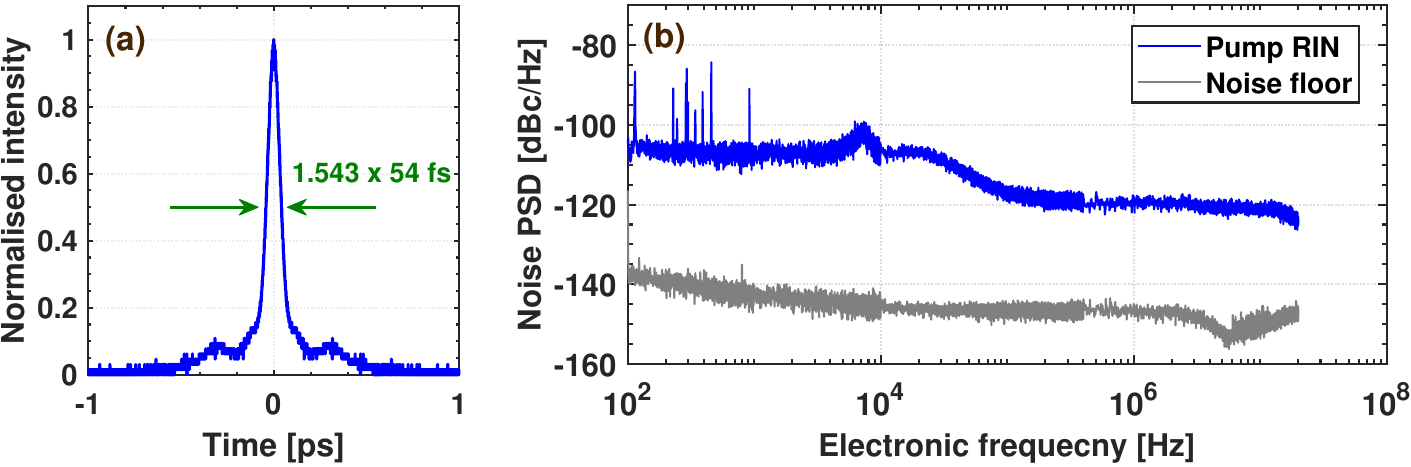}}
\caption{Experimentally measured characteristics of the 1.85~$\upmu$m output. (a) Intensity autocorrelation trace after the final compression fibre, corresponding to a 58~fs pulse duration assuming a Gaussian profile. (b) Frequency-dependent RIN of the output in blue and the detection system noise floor in grey.}
\label{Fig:AcRin}
\end{figure}
The pulse duration of the compressed 1.85~$\upmu$m output was characterised using an intensity autocorrelator (Femtochrome: FR-103MN). The measured autocorrelation trace, recorded after the final compression fibre, is shown in Fig.~\ref{Fig:AcRin}(a). The average output power was 210~mW. Assuming a Gaussian pulse profile, the FWHM duration was determined to be 58~fs. Fourier transformation of the measured output spectrum yields a transform-limited pulse duration of 43.3~fs, corresponding to a time-bandwidth product (TBP) of 0.433, while the measured pulse corresponds to a TBP of 0.58. This indicates residual higher-order dispersion (HOD) after compression not fully compensated by the PM2000D-based dispersion management~\cite{Pio18DispJb}. The pulse-to-pulse relative-intensity noise (RIN) of the 1.85~$\upmu$m output was measured using a photodiode (Thorlabs: PDA10D2) and an electronic spectrum analyser (ESA; Signal Hound: USB-SA44B), following the method in Ref.~\cite{Sco01EsaIe}. The photodiode signal was filtered using a 21~MHz low-pass filter (Crystek: CLPFL-0021-BNC), and a direct current (DC) block was used to remove frequency components below 7~Hz. 

The RIN was evaluated from 100~Hz to the system's maximum noise frequency of 20~MHz; the 100 Hz lower bound was chosen for reasonable acquisition time and does not significantly affect the integrated RIN, which is dominated by high-frequency noise. The background noise floor was measured with blocked optical input and converted to dBc/Hz using the same DC photocurrent as for the optical RIN trace. The measured frequency-dependent noise PSD of the 1.85~$\upmu$m output is shown in Fig.~\ref{Fig:AcRin}(b) in blue, with a measured RIN of 0.41\% across the electronic frequency range. For reference, the noise floor of the detection system is plotted in grey, corresponding to a RIN of 0.017\%. For comparison, the RIN of the seed before amplification was 0.20\%.

\begin{figure}[htbp!]
\centering
\fbox{\includegraphics[width=0.97\linewidth]{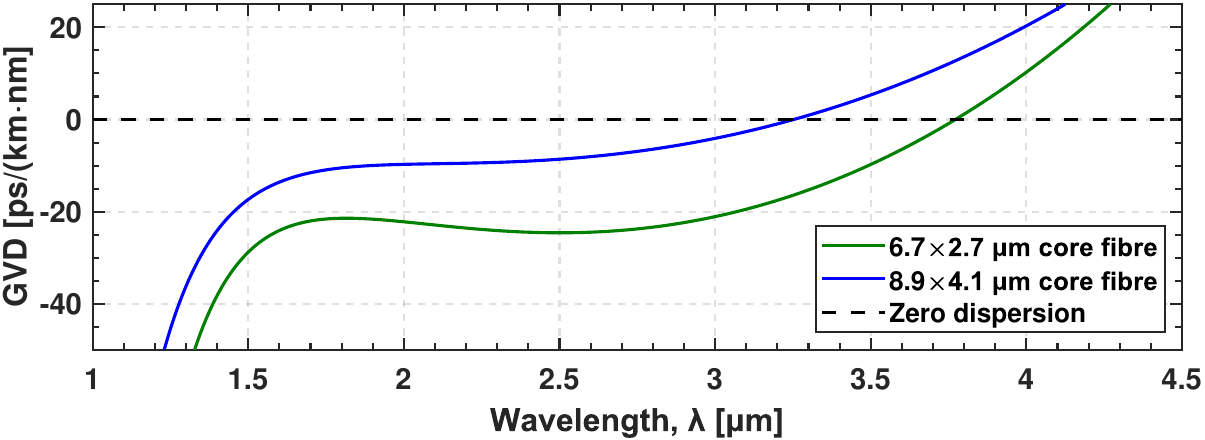}}
\caption{Measured GVD of one of the fundamental modes in PM ZBLAN fibres: 6.7$\times$2.7~$\upmu$m core in green; 8.9$\times$4.1~$\upmu$m core in blue; the dashed line indicates zero dispersion.}
\label{Fig:MeasDisp}
\end{figure}

To investigate ultra-low-noise SCG extending into the MIR, we identified two PM, normal-dispersion, elliptical-core ZBLAN fibres (FibreLabs: ZEF-2.7$\times$6.7$\mathrm{/}$95$\times$125-N and ZEF-4.5$\times$9$\mathrm{/}$125-N). Broadband group-velocity dispersion (GVD) and group-birefringence measurements of the two fundamental modes were performed using white-light interferometry~\cite{Hlu12DispJeos}. The fibres have core dimensions of 6.7$\times$2.7~$\upmu$m and 8.9$\times$4.1~$\upmu$m, and the measured GVD of one of the fundamental modes for both fibres is shown in Fig.~\ref{Fig:MeasDisp}. The 6.7$\times$2.7~$\upmu$m fibre exhibits normal dispersion up to 3.77~$\upmu$m, while the 8.9$\times$4.1~$\upmu$m fibre remains normal up to 3.25~$\upmu$m. These results confirm that the two elliptical-core PM ZBLAN fibres are suitable for generating ultra-low-noise SC up to 3.77~$\upmu$m and 3.25~$\upmu$m, respectively. Further details of the broadband GVD and group-birefringence measurements are provided in Ref.~\cite{Shr25EZbDSiD}.

The input and output ends of the PM ZBLAN fibres were cleaved using a Vytran cleaver (Thorlabs: LDC401A) with a tension of $\sim$100~g. For coupling, the bare-fibre facet output of the 58~fs, 1.85~$\upmu$m laser was mounted in a rotation mount (Thorlabs: HFR007) on a three-axis translation stage, and the ZBLAN input was likewise mounted on a three-axis translation stage. Coupling was achieved by physical contact, and the rotation angle was controlled to align the polarisation axes of both fibres. The polarisation extinction ratio of the pump is $\sim$20~dB, and the aligned all-PM architecture supports stable SC linear polarisation. The coupling efficiency was calculated as the ratio of the average power out of the ZBLAN fibre after an aspheric collimating lens (Thorlabs: C036TME-D) to the laser output power. SC was recorded using a 200~$\upmu$m-core ZBLAN collection fibre (Thorlabs: MZ22L1) connected to the OSA.
\begin{figure}[htbp!]
\centering
\fbox{\includegraphics[width=0.97\linewidth]{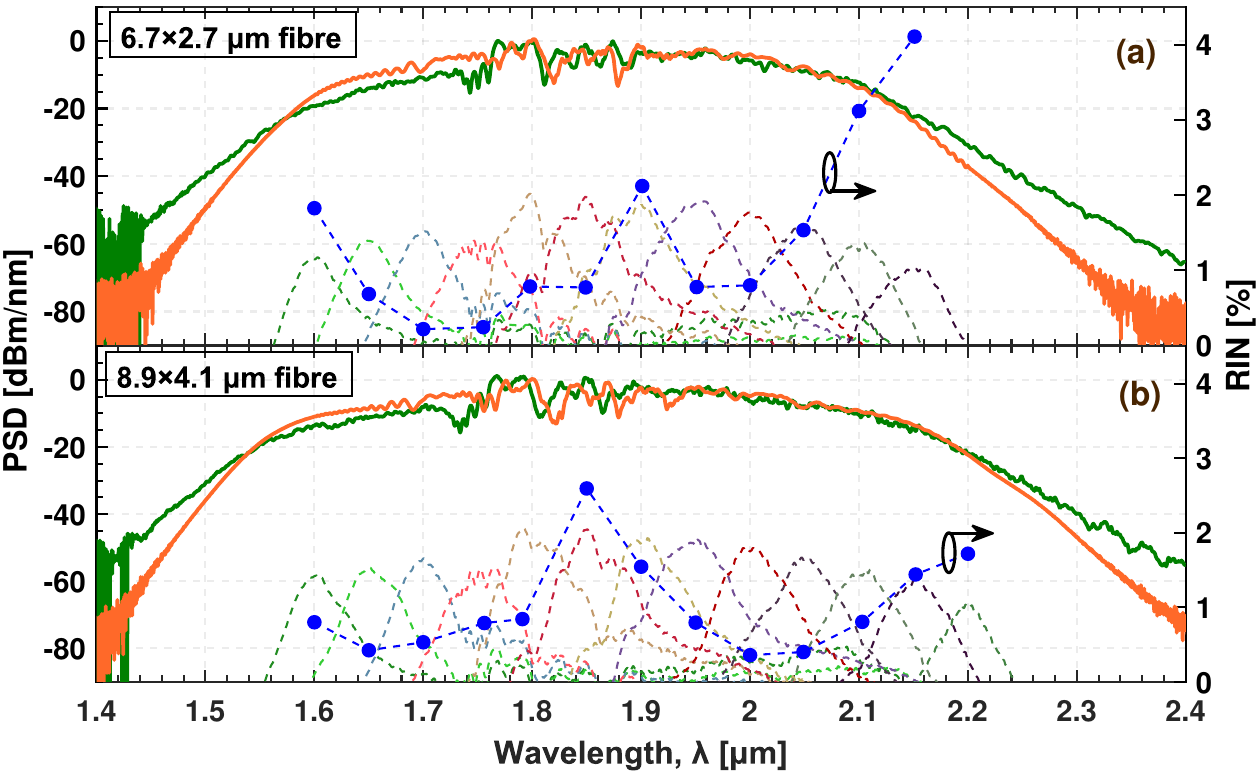}}
\caption{Experimentally measured SC in solid green, simulated SC in orange, and measured spectrally resolved RIN in blue for 6.7$\times$2.7~$\upmu$m (a) and 8.9$\times$4.1~$\upmu$m (b) core fibres. The peaks of the bandpass-filtered spectra are vertically offset by 45~dB relative to the SC; RIN refers to the right-hand y-axis.}
\label{Fig:LnSc}
\end{figure}

For the 6.7$\times$2.7~$\upmu$m core fibre, the fibre length was 1.55~m and the coupling efficiency was 61\%. The measured SC, when coupled to one of the fundamental modes of the fibre, is shown in Fig.~\ref{Fig:LnSc}(a) in solid green. The $-$30~dB spectral width is 660~nm, spanning 1.537--2.196~$\upmu$m, corresponding to 58 THz. Spectrally resolved pulse-to-pulse RIN was measured after bandpass filtering with a linear variable filter (LVF, Vortex Optical Coatings: LVF-1.3-2.6-3.5-15-0.5-2\%). The typical FWHM of each bandpass-filtered spectrum was $\sim$20~nm. The measured bandpass-filtered spectra are plotted in Fig.~\ref{Fig:LnSc}(a), and the corresponding measured RIN is shown in blue with respect to the right-hand y-axis. The RIN remains low across the band, reaching 0.22\% at 1.7~$\upmu$m. 

For the 8.9$\times$4.1~$\upmu$m core fibre, the fibre length was 2~m and the coupling efficiency achieved was 78\%. The SC out of the fibre, when coupled to one of the fundamental modes of the fibre, is shown in Fig.~\ref{Fig:LnSc}(b) in solid green. The --30~dB spectral width is 743~nm, spanning 1.507--2.250~$\upmu$m, corresponding to 66~THz.  The measured bandpass-filtered spectra are plotted in Fig.~\ref{Fig:LnSc}(b). The corresponding measured RIN is shown in blue with respect to the right-hand y-axis. The RIN of the SC out of this fibre also remains low across the band, reaching a minimum of 0.36\% at 2.0~$\upmu$m and increasing towards the spectral edges. For both fibres, the local RIN increase near the pump wavelength is attributed to spectral fine structure in the pump, which is transferred by the nonlinear broadening dynamics into locally enhanced intensity noise in the residual pump band. The somewhat higher RIN observed for the smaller-core fibre is attributed mainly to its narrower SC bandwidth, which was caused by lower input coupling efficiency. As a result, the long-wavelength measurement region lies closer to the spectral roll-off, where the RIN increases. Away from the spectral edge, the RIN is similar for both fibres. For clarity, all bandpass-filtered spectra shown in Fig.~\ref{Fig:LnSc} are vertically shifted by 45~dB relative to the SC.

To model the input pulse, the measured pump spectrum shown in orange in Fig.~\ref{Fig:IntSpec} was Fourier transformed to obtain its temporal profile. The transform-limited T$_{FWHM}$ was 43.3~fs. To match the autocorrelation trace in Fig.~\ref{Fig:AcRin}, the pulse was fitted by adding residual HOD, yielding third-order dispersion $\Phi_{3}=1.44\times10^{4}$~fs$^{3}$ and fourth-order dispersion $\Phi_{4}=3.45\times10^{6}$~fs$^{4}$. The effective area (A$_{eff}$) of each ZBLAN fibre was obtained from finite-element simulations. Scalar generalised nonlinear Schr\"{o}dinger equation simulations were then performed in the interaction picture~\cite{Hul07IpJt,Lae07MfdOe}. The simulation included the full measured GVD of ZBLAN fibres shown in Fig.~\ref{Fig:MeasDisp}, the wavelength-dependent A$_{eff}$, a double-Lorentzian Raman-gain fit to the measurement in Ref.~\cite{Chr11ZbRJb}, and the input pulse with HOD and quantum noise. The nonlinear refractive index used was n$_{2}=$2.55$\times$10$^{-20}$~m$^{2}\mathrm{/}$W, as in Ref.~\cite{Kub13ZbTJb}. At the pump wavelength, A$_{eff}$ was 21~$\upmu$m$^{2}$ for the 6.7$\times$2.7~$\upmu$m fibre and 33~$\upmu$m$^{2}$ for the 8.9$\times$4.1~$\upmu$m fibre; input peak powers were 47~kW and 53~kW, respectively. The simulated SC, shown in orange in Fig.~\ref{Fig:LnSc}(a) and~\ref{Fig:LnSc}(b), shows good agreement with the experimental results.

In conclusion, we have developed a fully PM, coherently seeded thulium-fibre CPA that delivers ultrafast, low-noise pump pulses at 1.85~$\upmu$m, extending earlier low-noise 2~$\upmu$m CPA work to a fully PM all-silica architecture~\cite{Hei202umSr,Ram19Opx}. By combining an Er:fibre-laser-driven ANDi seed with coherent amplification and compression in an all-fibre PM architecture, the system enables stable, alignment-free operation together with precise control of the pump pulse parameters. Using this source, we generated SC in normal-dispersion PM ZBLAN fibres spanning 1.5--2.25~$\upmu$m, with a minimum RIN of 0.22\% and an expected high degree of spectral coherence. The spectral broadening achieved in this work was primarily limited by the peak power available from our 1.85~$\upmu$m source. While the net spectral extension is therefore modest at the present power level, these results constitute, to the best of our knowledge, the first demonstration of ultra-low-noise SCG in a normal-dispersion fluoride fibre. Since the investigated PM ZBLAN fibres remain normally dispersive up to 3.77~$\upmu$m and 3.25~$\upmu$m, respectively, they provide a clear path towards further low-noise extension into the MIR, which is not accessible in conventional solid-core silica fibres.

\noindent\textbf{Funding.} {\small Danmarks Frie Forskningsfond project no. 2031-00009B; VILLUM Fonden (2021 Villum Investigator project no. 00037822: Table-Top Synchrotrons); Innovation Fund Denmark project no. 2105-00039B (HYPERSORT); the EU's Horizon Europe project no. 101058054 (TURBO); Schweizerischer Nationalfonds zur F\"{o}rderung der Wissenschaftlichen Forschung (TMPFP2\_210543, PCEFP2\_181222).}

\noindent\textbf{Acknowledgement.} {\small The authors thank FiberLabs Inc., Japan, for providing the ZBLAN fibre samples.}

\noindent\textbf{Disclosures.} {\small The authors declare no conflicts of interest.}

\noindent\textbf{Data Availability Statement.} {\small The data underlying the results presented in this paper are not publicly available at this time but may be obtained from the authors upon reasonable request.}
\bibliography{LnSOl25Final}
\end{document}